\documentclass[aps,prd,twocolumn,superscriptaddress]{revtex4-2}

\usepackage{amsmath}
\usepackage{epsfig}
\usepackage{cancel}
\usepackage{soul}
\usepackage{color, xcolor}
\usepackage[colorlinks=true,linkcolor=red,urlcolor=blue,citecolor=blue]{hyperref}
\begin{document}


\title{Purely gravitational dark matter production in warm inflation}


\author{Qing-Yang Wang}
\email[]{wangqy@ucas.ac.cn}
\affiliation{School of Fundamental Physics and Mathematical Sciences, Hangzhou Institute for Advanced Study, UCAS, Hangzhou 310024, China}
\author{Tianyu Jia}
\affiliation{Center for Gravitation and Cosmology, College of Physical Science and Technology, Yangzhou University, Yangzhou 225009, China}
\author{Pei-Ran Chen}
\affiliation{School of Fundamental Physics and Mathematical Sciences, Hangzhou Institute for Advanced Study, UCAS, Hangzhou 310024, China}
\affiliation{Institute of Theoretical Physics, Chinese Academy of Sciences, Beijing 100190, China}
\affiliation{University of Chinese Academy of Sciences, Beijing 100049, China}
\author{Yong Tang}
\affiliation{School of Fundamental Physics and Mathematical Sciences, Hangzhou Institute for Advanced Study, UCAS, Hangzhou 310024, China}
\affiliation{School of Astronomy and Space Science,
 University of Chinese Academy of Sciences (UCAS), Beijing 100049, China}
\affiliation{International Centre for Theoretical Physics Asia-Pacific, Beijing/Hangzhou, China}


\date{\today}

\begin{abstract}
We consider an appealing scenario for the production of purely gravitational dark matter in the background of warm inflation, a mechanism that maintains stable thermal bath during inflation. Through systematic investigation of various gravitational production channels, we reveal distinctive features compared to the standard inflation scenario. Notably, the inflaton annihilation channel in warm inflation exhibits markedly different thermodynamics from the standard inflation paradigm, leading to a suppression on the production of sub-inflaton-mass dark matter. For the production channel of inflationary vacuum fluctuations, we find an abundance-mass correlation of $\rho_\chi\propto m_\chi^{1/2}(m_\chi^{5/2})$ for the sub-Hubble-mass dark matter with minimal(conformal) coupling. Our results also indicate that a minimum temperature threshold of $10^{-6}M_P$ is necessary for warm inflation, which allows adequate dark matter production. With observational constraints, our results provide stringent limits on the mass range of purely gravitational dark matter with sufficient density: $10^{-8}-10^{-2}M_P$ for minimal coupling and $10^{-14}-10^{-2}M_P$ for conformal coupling.
\end{abstract}


\maketitle

\section{Introduction}
Dark matter (DM) is an essential component in the universe that revealed by a multitude of astronomical observations \cite{Zwicky:1933gu,Smith:1936mlg,Rubin:1970zza,Clowe:2006eq,Planck:2018vyg,Bertone:2016nfn,Cirelli:2024ssz}. The density fraction of it may exceed five times that of visible matter, yet it remains undetected by direct experimental means \cite{Schumann:2019eaa,Lin:2019uvt,Misiaszek:2023sxe}, and has only been identified indirectly through its substantial gravitational influence. Growing experimental constraints reflect the extreme weakness of direct interaction between DM and the standard model (SM) particles, which constitute visible matter. It is even possible that there is no other interaction between DM and visible particles besides gravity. This is the simplest possibility of DM, which is called the purely gravitational DM \cite{Tang:2016vch,Ema:2018ucl,Ema:2019yrd,Mambrini:2021zpp,Clery:2021bwz,Henrich:2024rux,Hashiba:2018tbu,Fairbairn:2018bsw,Carney:2019pza,Redi:2020ffc,Haque:2021mab,Garcia:2023qab}.

Production of purely gravitational DM generally occurs in the very early universe, e.g., during inflation and following reheating stage. Two feasible production mechanisms are usually discussed. One is called the cosmological gravitational particle production (CGPP) \cite{Kolb:2023ydq,Parker:1968mv,Parker:1969au,Parker:1971pt,Ford:1986sy,Lyth:1996yj,Kuzmin:1998kk,Chung:1998zb,Kolb:1998ki,Chung:2001cb,Li:2019ves,Ema:2018ucl,Ema:2019yrd,Gorbunov:2012ij,Markkanen:2015xuw,Ema:2016hlw,Alonso-Alvarez:2018tus,Cembranos:2019qlm,Chung:2011ck,Graham:2015rva,Ahmed:2020fhc,Kolb:2020fwh,Gross:2020zam,Cembranos:2023qph,Ozsoy:2023gnl,Kallosh:1999jj,Lyth:1999yc,Kolb:2021xfn,Kolb:2023dzp,Wang:2022ojc,Racco:2024aac,Gorji:2025tos}, which refers to the particle production from the vacuum of an expanding universe. This mechanism arises from the temporal variation in the vacuum state of the quantum field in the cosmological background. The vacuum state at the previous moment can evolve into an excited state relative to the vacuum state at the next moment, indicating the particle production from the vacuum. There is a detailed review article discussed the CGPP of DM \cite{Kolb:2023ydq}. Another mechanism is to connect the thermal bath and DM through graviton as a medium, which is also known as the gravitational DM freeze-in \cite{Hall:2009bx,Tang:2017hvq,Kolb:2017jvz,Hashiba:2018tbu,Garcia:2020wiy,Henrich:2024rux,Bernal:2025fdr,Mambrini:2021zpp,Clery:2021bwz,deSouza:2024oaz}. Efficiency of this mechanism strongly depends on the temperature of the thermal bath. In the scenario of reheating, it necessitates an exceptionally efficient reheating mechanism to achieve a sufficiently high temperature for DM freeze-in.

Reheating is a conventional scenario that connects the inflationary phase and the radiation-dominant (RD) phase in the standard inflation (SI) theory \cite{Albrecht:1982mp,Kofman:1994rk,Kofman:1997yn,Shtanov:1994ce,Allahverdi:2010xz,Amin:2014eta}. However, it is not the only possibility. An alternative scenario suggests that the inflaton field may have a strong interaction with radiation during inflation, thereby dissipating energy into radiation to sustain a stable thermal bath, up until the universe completes the transition from inflation to the RD stage. It is known as the warm inflation (WI) paradigm \cite{Berera:1995ie,Berera:1996nv,Berera:1996fm,Berera:1998px,Berera:1999ws,Taylor:2000ze,Hall:2003zp,Berera:2008ar,Bastero-Gil:2009sdq,Cai:2010wt,Bartrum:2013fia,Benetti:2016jhf,Bastero-Gil:2016qru,Bastero-Gil:2017yzb,Motaharfar:2018zyb,Arya:2019wck,Berghaus:2019whh,Brandenberger:2020oav,Das:2020lut,Bastero-Gil:2021fac,Montefalcone:2022jfw,Montefalcone:2023pvh,Kamali:2023lzq,Berera:2023liv,Santos:2024pix,Santos:2024plu,Berghaus:2025dqi}. There are many motivations for WI. For instance, prior studies indicate that WI with strong dissipation can inhibit super-Planckian field variation for the inflaton \cite{Berera:2003yyp}, thereby facilitating the realization of small-field inflation theory. This is a nice feature from the perspective of effective field theory. Furthermore, WI is capable of producing an enhanced scalar perturbation spectrum, which consequently suppresses the tensor-to-scalar ratio of the primordial perturbations \cite{Bartrum:2013fia,Benetti:2016jhf}. As a result, it revives many inflationary models previously deemed unfavorable, such as $\phi^2$ and $\phi^4$ potentials, which have been excluded in the SI paradigm by Planck and BICEP/$Keck$ results \cite{Planck:2018jri,BICEP:2021xfz}.

In contrast to SI with reheating, WI is more likely to attain a higher radiation temperature due to the absence of interval between the inflationary and RD stages, thus it enhances the likelihood of producing massive DM particles via freeze-in mechanism. Recent findings in \cite{Freese:2024ogj} demonstrate that freeze-in production of DM in WI is more efficient than that in SI. The yield is highly sensitive on the mass dimensions of the non-renormalizable interaction operator. If the mass dimension is large enough, the DM yield will increase by many orders of magnitude compared to that of SI with reheating. 

We are curious about the gravitational freeze-in mechanism of DM in the WI background. Considering the gravitational annihilation of SM particles and inflatons, this study demonstrates the dependence of DM yield on the WI temperature and DM mass, indicating a suppression on the production of sub-inflaton-mass DM. The contribution of CGPP mechanism is also taken into account, revealing different abundance-mass correlations from SI. In addition, the combined observational constraints to the parameter space will also be presented, showing stringent limits on the mass range of DM.

This paper is organized as follows. In Sec.~II, we start with the background dynamics of WI, highlighting its basic features. Then in Sec.~III, we calculate the cross-section of graviton-mediated annihilation and derive the DM yield with the Boltzmann equation. Later in Sec.~IV, we investigate the DM relic abundance for three different channels. The observational constraints on the parameter space is discussed in Sec.~V. And the conclusions are given in the final section. 

The following conventions are adopted: FLRW metric $g_{\mu\nu}x^\mu x^\nu=-\mathrm dt^2+a^2(t)\mathrm d\vec x^2$, natural unit $\hbar=c=1$, and Planck mass $M_P\equiv1/\sqrt{8\pi G}=2.435\times10^{18}\mathrm{GeV}=1$.

\section{Background dynamics of warm inflation}
During WI, the background dynamics of the universe are collectively governed by both inflaton and radiation, with a persistent energy transfer from the inflaton to radiation, which ensures their energy densities gradually converging as inflation progresses. We commence our analysis with the energy conversion equations pertinent to each component of the universe:
\begin{align}
    \dot\rho_\alpha+3H\left(\rho_\alpha+p_\alpha\right)=\mathcal E_\alpha,
\end{align}
where $H\equiv\dot a/a$ is the Hubble parameter, $\rho$ and $p$ are the energy density and the pressure of each component $\alpha$, respectively. $\mathcal E_\alpha$ denotes the energy transfer between various components. The energy conversion requires $\sum_\alpha\mathcal E_\alpha=0$. For WI, this set of equations can be written as:
\begin{gather}
    \dot\rho_\phi+3H\left(\rho_\phi+p_\phi\right)=-\Upsilon\left(\rho_\phi+p_\phi\right),\label{rhophi1}\\
    \dot\rho_r+3H\left(\rho_r+p_r\right)=\Upsilon\left(\rho_\phi+p_\phi\right),\label{rhor1}
\end{gather}
where the subscript $\phi$ denotes inflaton and $r$ denotes radiation. $\Upsilon$ is a dissipation coefficient that measures the efficiency of energy transfer. It is usually a function of the value of inflaton and the radiation temperature.

Generally, for a homogeneous inflaton field with potential $V(\phi)$, the energy density and the pressure can be written as:
\begin{gather}
    \rho_\phi=\frac{1}{2}\dot\phi^2+V(\phi),\\
    p_\phi=\frac{1}{2}\dot\phi^2-V(\phi).
\end{gather}
Substituting them into Eqs.~(\ref{rhophi1}) and (\ref{rhor1}), and introducing the equation of state for radiation, $p_r=\rho_r/3$, we derive the equation of motion for inflaton and the energy evolution equation for radiation:
\begin{gather}
    \ddot\phi+\left(3H+\Upsilon\right)\dot\phi+\frac{\mathrm dV}{\mathrm d\phi}=0,\label{phieq}\\
    \dot\rho_r+4H\rho_r=\Upsilon\dot\phi^2.\label{rhoreq}
\end{gather}
These two equations, together with the Friedmann equation in a flat universe:
\begin{align}\label{fldm}
    H^2=\frac{1}{3}\left(\rho_\phi+\rho_r\right),
\end{align}
make up the background dynamics of WI.

It is clear that the impact of dissipation on the inflaton field is analogous to that of Hubble friction, as it reduces the field's evolution rate $\dot\phi$ and diminishes its kinetic energy. Given that $\mathrm d\phi/\mathrm dN=\dot\phi/H$ ($N\equiv\ln a$ is the $e$-folding number of inflation), a decrease in $\dot\phi$ corresponds to a reduced change in $\phi$ for the same $e$-folding expansion. That's why WI with strong dissipation is capable of preventing super-Planckian field variations. For convenience, a dimensionless dissipation coefficient, $Q\equiv\Upsilon/(3H)$, is introduced, which allows for a clear comparison between the roles of dissipation and Hubble friction concerning the inflaton field's kinetic energy. When $Q\ll1$, the WI is in the weak dissipative regime, where dissipation plays a minor role. Conversely, when $Q>1$, it is in the strong dissipative regime, wherein dissipation significantly influences the evolution of inflaton field and the properties of primordial perturbations. Of course, $Q$ is a time-varying parameter. WI may transfer from the weak dissipative regime to the the strong dissipative regime, or vice versa, with the progress of inflation.

Generally, the specific model that can realize WI is rather complex. However, in the context of DM production processes under examination, the outcomes primarily hinge on the late-stage behavior of WI. Therefore, we can afford to make certain simplifications concerning the model exclusively at the terminal phase of WI. First, we consider an inflationary potential proportional to $\phi^2$ at its minimum:
\begin{align}\label{V}
    V(\phi)=\frac{1}{2}m_\phi^2\phi^2.
\end{align}
Numerous inflation models justify such assumptions at the bottom of the potential, including the Starobinsky $R^2$ model and its higher-order extensions \cite{Starobinsky:1980te,Huang:2013hsb,Asaka:2015vza,Cheong:2020rao,Rodrigues-da-Silva:2021jab,Ivanov:2021chn,Shtanov:2022pdx,Wang:2023hsb}, the Higgs inflationary model \cite{Bezrukov:2007ep}, $n=1$ $\alpha$-attractor models \cite{Kallosh:2013hoa,Kallosh:2013yoa}, etc. Then we assume that the dissipation coefficient $\Upsilon$ exhibits a power law in relation to the value of inflaton $\phi$ and the radiation temperature $T$:
\begin{align}\label{Upsl}
    \Upsilon(\phi,T)=C_\Upsilon T^c\phi^p M_P^{1-p-c},
\end{align}
where $C_\Upsilon$ is a constant, and $c$, $p$ are numerical powers. This assumption is quite common in WI researches \cite{Benetti:2016jhf,Das:2020lut,Kamali:2023lzq}. Many models can derive a dissipation coefficient in this form.

\begin{figure}[!t]
    \centering\includegraphics[width=0.49\textwidth]{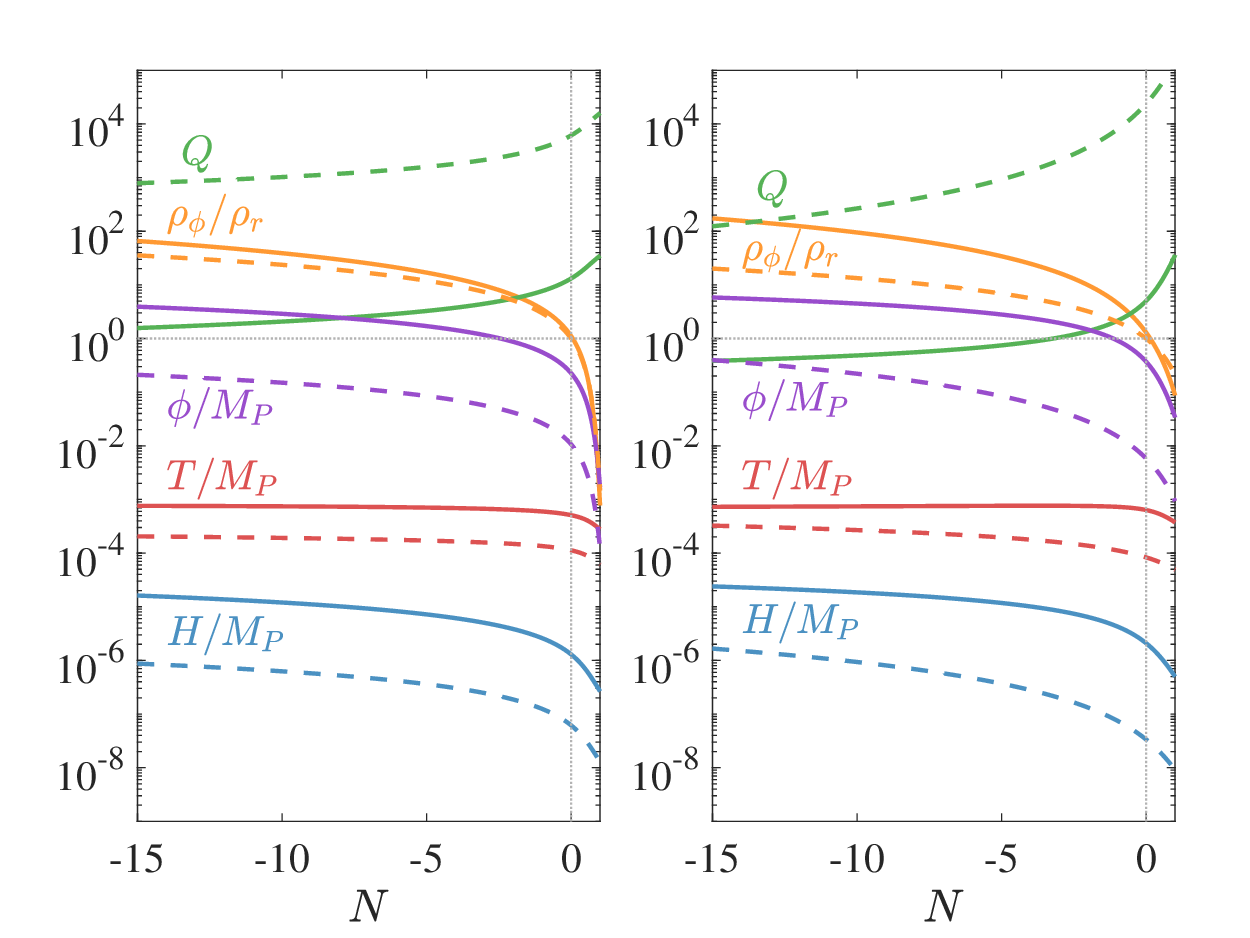}
    \caption{Evolution of various dimensionless quantities during the late stage of WI. The inflationary potential is assumed to possess a $\phi^2$ bottom with $m_\phi=10^{-5}M_P$. In the left panel we set $p=0$, $c=1$, and $C_\Upsilon=0.1(10)$ for the solid(dashed) curves. In the right panel we set $p=0$, $c=-1$, and $C_\Upsilon=2\times10^{-8}(2\times10^{-7})$ for the solid(dashed) curves. The end of WI is fixed at $N=0$.\label{fig1}}
\end{figure}

Substituting Eqs.~(\ref{V}), (\ref{Upsl}) and the expression of temperature:
\begin{align}
    T=\left(\frac{30\rho_r}{\pi^2 g_*}\right)^{1/4},
\end{align}
into Eqs.~(\ref{phieq}), (\ref{rhoreq}), and (\ref{fldm}), the background dynamics of WI can be numerically solved. Let us now choose several sets of parameters to demonstrate the evolution of various physical quantities during the late stage of WI. The results are depicted in Fig.~\ref{fig1}. Here we fix $m_\phi=10^{-5}M_P$, $p=0$, and set $c=1(-1)$ for the left(right) panel of the figure. The solid and dashed curves signify different values of $C_\Upsilon$, or different initial value of $Q$, respectively. It is evident that the radiation sustains an almost constant temperature with $T\gg H$ until the end of inflation. This is an important characteristic of WI that will be used in this study. Note that in the WI scenario, the end of inflation is driven by the energy density of radiation $\rho_r$ surpassing that of the inflaton $\rho_\phi$, rendering the $\epsilon_V$ criterion used in the SI framework inapplicable. Only the $\epsilon_H$ criterion is valid in WI, that is, WI ends when $\epsilon_H\equiv-\dot H/H^2>1$, or equivalently, $\ddot a<0$.

\section{Gravitational dark matter production}
Now we consider the production of purely gravitational DM during WI. We start with the following action:
\begin{align}
    S=\int\mathrm dx^4\sqrt{-g}\left[\frac{M_P^2}{2}R+\mathcal L_M\right],
\end{align}
where $\mathcal L_M$ refers to all material fields, including inflaton, radiation, and DM. Gravity plays a complex role in this system and poses challenges when addressed through non-perturbative methods. Nonetheless, if the gravitational perturbation is weak enough, the gravity induced process in this system can be effectively analyzed with the effective field theory approach.

To construct an effective field theory of gravity, we decompose the metric into a homogeneous FLRW background $\tilde g_{\mu\nu}$ and a minor perturbation $h_{\mu\nu}$:
\begin{align}
    g_{\mu\nu}=\tilde g_{\mu\nu}+\frac{2}{M_P}h_{\mu\nu}.
\end{align}
Then the action we consider can be expanded as:
\begin{align}\label{S1}
    S=&\int\mathrm dx^4\Big\{\tilde{\mathcal L}_M+\mathcal L_2(h)\nonumber\\
    &+\frac{1}{M_P}\left[h_{\mu\nu}T^{\mu\nu}_{M}+\mathcal L_3(h)\right]+...\Big\}.
\end{align}
Here, $\tilde{\mathcal L}_M$ refers to the material fields in the FLRW background, $\mathcal L_n(h)$ refers to the $n$-th order term of $h_{\mu\nu}$, and $T^{\mu\nu}_{M}$ is the energy-momentum tensor of material fields. If we consider the perturbative quantization of gravity, then $h_{\mu\nu}$ can be regarded as graviton, and $h_{\mu\nu}T^{\mu\nu}_{M}$ describes the leading order interaction between graviton and matter.

\begin{figure}[!t]
    \centering\includegraphics[width=0.4\textwidth]{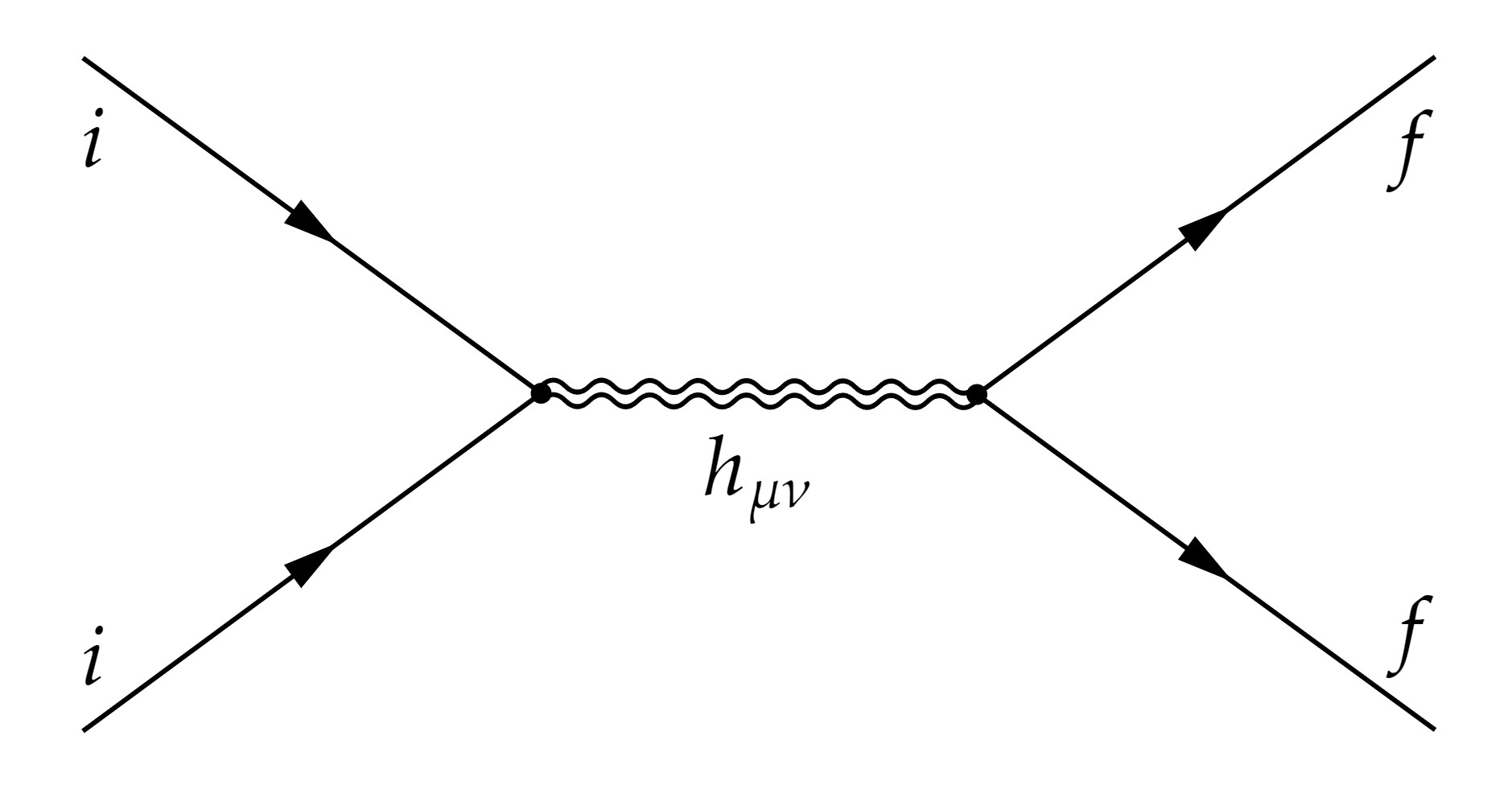}
    \caption{Illustration of the $s$-channel graviton-mediated annihilation process. The initial and final state are labeled by $i$ and $f$. Arrows mean the directions of momenta.\label{fig2}}
\end{figure}

The action Eq.~(\ref{S1}) actually divides DM production into two channels. One is the CGPP mechanism mentioned earlier, which is induced by the DM part in $\tilde{\mathcal L}_M$. The other is the freeze-in mechanism through the graviton-mediated interactions between DM and other particles, induced by $h_{\mu\nu}T^{\mu\nu}_{M}$ terms. The interaction here is mainly the $s$-channel annihilation process, as shown in Fig.~\ref{fig2}. During WI, both radiation particles and inflatons possess considerable energy due to the high temperature, which significantly enhances the gravitational scattering cross-section, thereby facilitating the production of massive DM particles.

We first focus on the DM production via graviton-mediated annihilation. This process entails a two-step calculation. The first step requires the computation of the annihilation cross-section. Following this, the Boltzmann equation is solved using the calculated cross-section to determine the relic abundance of DM.

\subsection{Cross-section of graviton-mediated annihilation}
To calculate the cross-section, we first give the specific form of the energy-momentum tensor $T^{\mu\nu}_{M}$. In principle, the energy-momentum tensor here should be considered under the FLRW metric. However, as we are studying the scattering between particles, the temporal and spatial scales of these processes are very short, we can ignore the influence of curvature or cosmic expansion on each scattering process and only consider the energy-momentum tensor under the Minkowski metric $\eta_{\mu\nu}$. For scalar $\varphi$, Dirac fermion $\psi$, and massive vector $A_\mu$, the energy-momentum tensors are:
\begin{align}
    T^{\mu\nu}_{\varphi}=&~\partial^\mu\varphi\partial^\nu\varphi-\frac{1}{2}\eta^{\mu\nu}\partial^\alpha\varphi\partial_\alpha\varphi+\frac{1}{2}\eta^{\mu\nu}m_\varphi^2\varphi^2,\\
    T^{\mu\nu}_{\psi}=&-\eta^{\mu\nu}\left(\bar{\psi}i\cancel\partial\psi-m_\psi\bar{\psi}\psi\right)\nonumber\\
    &+\frac{1}{2}\bar{\psi}i\gamma^{\mu}\partial^{\nu}\psi+\frac{1}{2}\bar{\psi}i\gamma^{\nu}\partial^{\mu}\psi+\frac{1}{2}\eta^{\mu\nu}\partial^{\alpha}\left(\bar{\psi}i\gamma_{\alpha}\psi\right)\nonumber\\
    &-\frac{1}{4} \partial^{\mu}\left(\bar{\psi}i\gamma^{\nu}\psi\right)-\frac{1}{4}\partial^{\nu}\left(\bar{\psi}i\gamma^{\mu}\psi\right),\\
    T^{\mu\nu}_{A}=&~\frac{1}{4}\eta^{\mu\nu}F^{\alpha\beta}F_{\alpha\beta}-F^{\mu\alpha}F_{\alpha}^{\nu}-\frac{1}{2}\eta^{\mu\nu}m_{A}^2 A^{\alpha} A_{\alpha}\nonumber\\
    &+m_{A}^2 A^{\mu}A^{\nu}.
\end{align}

The matrix element $\mathcal M$ of the annihilation process shown as Fig.~\ref{fig2} can be calculated with these energy-momentum tensors and the propagator of graviton with momentum $p$ in the harmonic gauge:
\begin{align}
    D_{\mu\nu,\rho\sigma}(p)=\frac{\eta_{\mu\rho}\eta_{\nu\sigma}+\eta_{\mu\sigma}\eta_{\nu\rho}-\eta_{\mu\nu}\eta_{\rho\sigma}}{2p^2}.
\end{align}
For convenience, the computed results are expressed as the following integral:
\begin{align}
    \mathcal A\equiv\int\mathrm d\cos\theta\sum_\mathrm{pol}|\mathcal M|^2.
\end{align}
We consider the DM as a minimally coupled real scalar $\chi$ with mass $m_\chi$, and treat the initial states as scalar $\varphi$, fermion $\psi$, massive vector $A_\mu$, and massless vector $\gamma$ respectively. Then we derive:
\begin{align}
    \mathcal A_{\varphi\rightarrow\chi}=&~\frac{1}{M_P^4}\left[\frac{56m_\varphi^4m_\chi^4}{15s^2}-\frac{8m_\varphi^2m_\chi^2}{15 s}\left(m_\varphi^2+m_\chi^2\right)\right.\nonumber\\
    &\left.+\frac{2}{5}\left(m_\varphi^4+4m_\varphi^2m_\chi^2+m_\chi^4\right)\right.\nonumber\\
    &\left.+\frac{2s}{15}\left(m_\varphi^2+m_\chi^2\right)+\frac{s^2}{15}\right],\\
    \mathcal A_{\psi\rightarrow\chi}=&~\frac{1}{M_P^4}\left[-\frac{112 m_\psi^4m_\chi^4}{15s^2}-\frac{4m_\psi^2m_\chi^2}{15s}\left(m_\chi^2-4 m_\psi^2\right)\right.\nonumber\\
    &\left.+\frac{4}{15}\left(2m_\chi^4+3m_\psi^2m_\chi^2-3m_\psi^4\right)\right.\nonumber\\
    &\left.+\frac{s}{15}\left(4m_\chi^2-m_\psi^2\right)+\frac{s^2}{30}\right],\\
    \mathcal A_{A\rightarrow\chi}=&~\frac{1}{M_P^4}\left[\frac{808 m_A^4m_\chi^4}{15s^2}-\frac{8m_A^2m_\chi^2}{5s}\left(11m_\chi^2+ m_A^2\right)\right.\nonumber\\
    &\left.+\frac{2}{15}\left(19m_\chi^4+76m_A^2m_\chi^2+49m_A^4\right)\right.\nonumber\\
    &\left.-\frac{14s}{15}\left(m_\chi^2+m_A^2\right)+\frac{s^2}{5}\right],\label{AA}\\
    \mathcal A_{\gamma\rightarrow\chi}=&~\frac{1}{M_P^4}\left[\frac{2}{15}\left(s-4m_\chi^2\right)^2\right],\label{Agamma}
\end{align}
here $s\equiv\left(p_1+p_2\right)^2=\left(p_3+p_4\right)^2$. More general calculation results can be found in \cite{Tang:2017hvq}. Note that the result of massless vector Eq.~(\ref{Agamma}) cannot be regarded as the $m_A\rightarrow0$ limit of Eq.~(\ref{AA}).

Finally, the cross-section can be written as:
\begin{align}
    \sigma=\frac{1}{32\pi s\left(\mathcal Sg_i^2\right)}\frac{|\vec p_f|}{|\vec p_i|}\mathcal A,
\end{align}
where $\mathcal S$ is a symmetric factor. If the final particles are identical, then $\mathcal S=2$, else $\mathcal S=1$. $g_i$ is the degrees of freedom for initial state. For the annihilation of radiation particles into DM, if there are only SM particles in the thermal bath and the temperature exceeds the electroweak scale, making the SM particles all massless, we have:
\begin{align}\label{ASM}
    \mathcal A_{\mathrm{SM}\rightarrow\chi}=2\mathcal A_{\varphi\rightarrow\chi}+\frac{45}{2}\mathcal A_{\psi\rightarrow\chi}+12\mathcal A_{\gamma\rightarrow\chi},
\end{align}
here the factor of $1/2$ is considered for neutrinos since they are Weyl particles.

\subsection{Dark matter yield}
For $1+2\leftrightarrow3+4$ thermal scattering process in an expanding universe, the number density $n$ of final particles obeys the following Boltzmann equation:
\begin{align}
    \frac{1}{a^3}\frac{\mathrm d}{\mathrm dt}\left(na^3\right)=\mathcal R(T),
\end{align}
where $\mathcal R(T)$, we call it the interaction term, is a function of the cosmic temperature $T$, which describes the production rate of final particles. For the process we consider, the distribution of initial particles can be approximated as the Maxwell–Boltzmann distribution, then we have the interaction term with the following form \cite{Tang:2017hvq}:
\begin{align}\label{RT}
    \mathcal R(T)=\frac{Tg_i^2}{32\pi^4}\int\mathrm ds\sigma\sqrt{s}\left(s-4m_i^2\right)K_1\left(\frac{\sqrt{s}}{T}\right),
\end{align}
where $m_i=m_1=m_2$ is the mass of initial particles, $K_1$ is the modified Bessel function of the second kind with order one. The derivation of this equation is shown in the appendix.

Now we discuss the DM yield $Y_\chi$. It is defined as the ratio of the DM number density $n_\chi$ and the entropy density $\mathbf s$ of thermal bath, $Y_\chi\equiv n_\chi/\mathbf s$, where
\begin{align}
    \mathbf s=\frac{2\pi^2}{45}g_{*,S}T^3.
\end{align}
From the Boltzmann equation, the yield over a period of time is derived as:
\begin{align}
    Y_\chi(t_0)=\frac{45a_0^{-3}}{2\pi^2g_{*,S}T_0^3}\int_0^{t_0}a^3\mathcal R\mathrm dt,
\end{align}
or replacing the variable as the cosmic scale:
\begin{align}\label{Y}
    Y_\chi(a_0)=\frac{45a_0^{-3}}{2\pi^2g_{*,S}T_0^3}\int_0^{a_0} \frac{a^2\mathcal R}{H}\mathrm da.
\end{align}
We will prove later that $\mathcal R\propto T^8$ for the gravitational freeze-in process we study. Given that $T$ and $H$ are almost constant before the end of WI, and $T\propto a^{-1}$, $H\propto a^{-2}$ in the RD stage after WI, we can easily notice that the integrand $\mathcal I\equiv a^2\mathcal R/H$ has a peak at the end of WI. Therefore, the DM yield is principally determined by the integral calculated near the end of WI, which is almost solely dependent on the radiation temperature at the end of WI (denoted by $T_e$).

\begin{figure}[!t]
    \centering\includegraphics[width=0.49\textwidth]{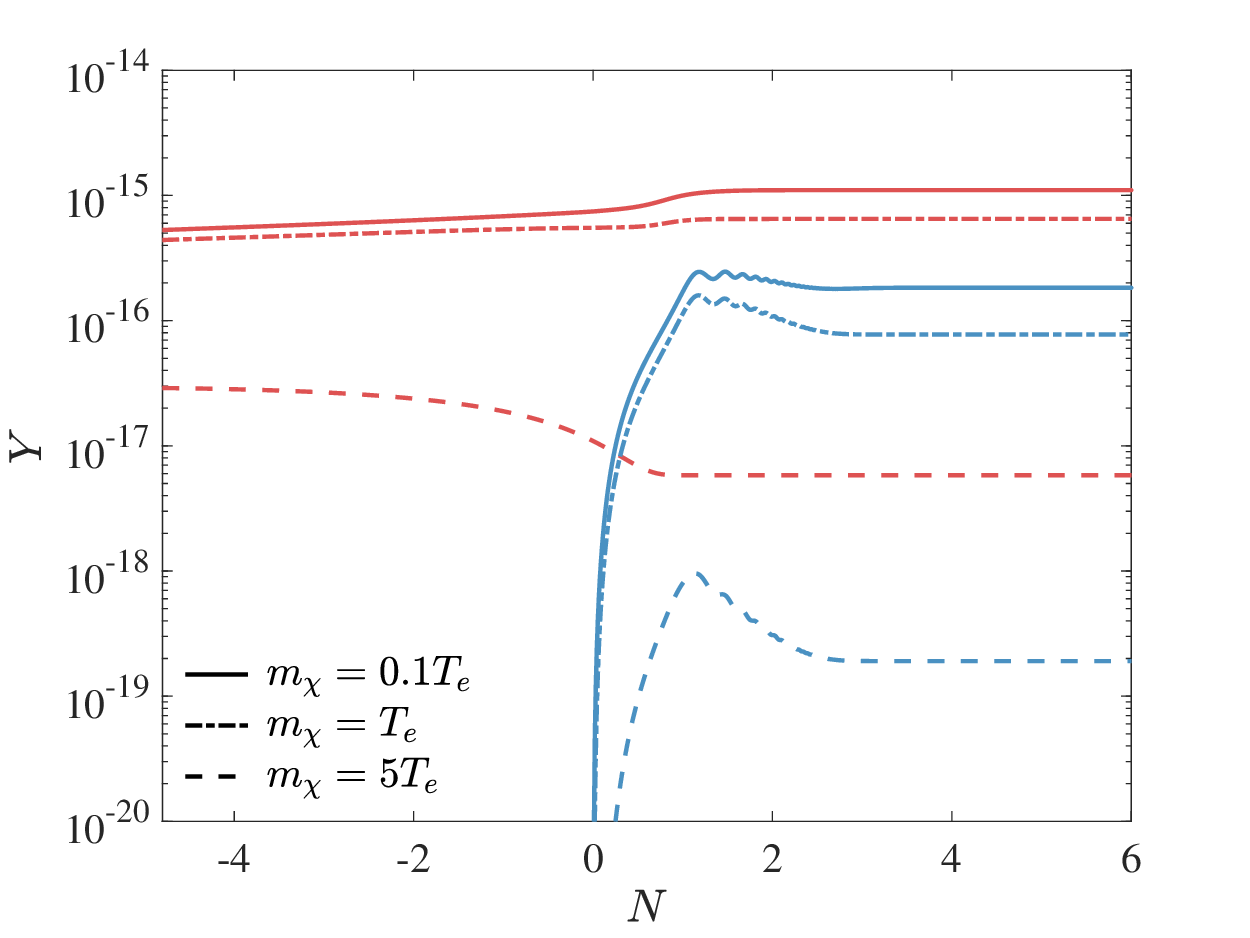}
    \caption{DM yield via gravitational annihilation during and after WI (red curves) and SI with reheating (blue curves). The inflationary potential is assumed to possess a $\phi^2$ bottom with $m_\phi=10^{-5}M_P$. For WI, we set $p=0$, $c=1$, and $C_\Upsilon=0.1$. The radiation temperature is $T_e=5\times10^{-4}M_P$ at the end of WI. For SI, we set the reheating efficiency $\Gamma=0.1H_e=6.5\times10^{-7}M_P$. The max temperature during reheating is $T_\mathrm{max}=5\times10^{-4}M_P$. The initial state is set as the average of all SM particles, and the DM is considered as a real scalar with mass of $m_\chi=0.1T_e,~T_e,~5T_e$ respectively. The end of WI is fixed at $N=0$.\label{fig3}}
\end{figure}

We show some calculation results for scalar DM yield in Fig.~\ref{fig3}. The red curves are for a kind of WI model with $T_e=5\times10^{-4}M_P$. The initial state here is considered as the average of all SM particles, which means the result in Eq.~(\ref{ASM}) is adopted. The interaction term in this case has a analytical expression:
\begin{align}\label{RSM}
    \mathcal R_{\mathrm{SM}\rightarrow\chi}=\frac{149}{160\pi^5}\frac{T^8}{M_P^4},
\end{align}
for $m_i^2/s\rightarrow0$ and $m_\chi^2/s\rightarrow0$. It is observed in Fig.~\ref{fig3} that the DM achieves stable yield during the WI period. This characteristic is understandable, because $\mathcal R$ is almost constant during WI due to the approximate invariance of temperature. Then when inflation ends, the yield undergoes certain changes and subsequently reaches a stable value again. This is actually the relic abundance of DM today. In the case we consider, $m_\chi=T_e$ serves as a watershed for the DM yield. If $m_\chi\ll T_e$, $Y$ becomes independent of $m_\chi$. It evolves as the red solid curve in Fig.~\ref{fig3}. As $m_\chi$ approaches $T_e$, there is a reduction in the yield $Y$ relative to the case of $m_\chi \ll T_e$. When $m_\chi$ exceeds $T_e$, the impact becomes more pronounced. The yield is not only apparently low during the WI period, but also further slumps when inflation ends. This is explicit for the scarcity of radiation particles with enough energy to produce DM with mass far larger than $T_e$.

For comparison, the blue curves in Fig.~\ref{fig3} show the DM yield for SI with max temperature $T_\mathrm{max}=5\times10^{-4}M_P$ during reheating and $T_\mathrm{reh}=3\times10^{-4}M_P$ at the end of reheating. Here we adopt the same $\phi^2$-bottom potential as WI. Generally, in contrast with the WI models, most SI models with reheating have a lower DM yield for the same potential and DM candidate. This discrepancy arises from the difficulties they encounter in attaining temperatures as high as that in the WI scenario. Unless the reheating efficiency $\Gamma$ exceeds the Hubble parameter $H_e$ at the end of inflation, the integrand of Eq.~(\ref{Y}) in the reheating scenario represents merely the tail of that in the WI scenario due to the postponed formation of thermal bath.

\section{Relic abundance of dark matter}
Now we investigate the relationship between the relic abundance of DM and relative parameters, including DM mass $m_\chi$ and the radiation temperature $T_e$ at the end of WI. The relic abundance is represented by the present-day energy density of DM:
\begin{align}
    \rho_{\chi,0}=m_\chi \mathbf s_0 Y_{t\rightarrow\infty}.
\end{align}
The observation of Planck Collaboration \cite{Planck:2018vyg} gives $\rho_{\chi,0}^{\mathrm{obs}}=1.24~\mathrm{GeV/m^3}=2.73\times10^{-121}M_P^4$.

\subsection{Annihilation of standard model particles}
We first focus on the SM particles annihilating into scalar DM. In the WI scenario, the SM particles in the thermal bath are expected to obtain a thermal mass arising from the coupling with inflaton field. Nonetheless, the thermal mass is usually much smaller than $T_e$, hence we still regard the thermal bath as composed of massless particles. With some simple approximations, the relic abundance can be analytically calculated for the $m_\chi\ll T_e$ case. We assume:
\begin{align}
    T\simeq\begin{cases}
    T_e & \text{for } a\leq a_e, \\
    T_ea_e/a & \text{for } a>a_e,
\end{cases}
\end{align}
\begin{align}
    \rho\simeq2\rho_r~~~\text{for } a\sim a_e,
\end{align}
for general WI models, where $a_e$ denotes the end of WI. Substituting them and Eq.~(\ref{RSM}) into Eq.~(\ref{Y}), we derive the final yield:
\begin{align}
    Y_{t\rightarrow\infty}=\frac{1341\sqrt{5}}{32\pi^8g_*^{3/2}}\frac{T_e^3}{M_P^3}=8.95\times10^{-6}\frac{T_e^3}{M_P^3},
\end{align}
where $g_*=106.75$ is taken for SM particles. Then the relic abundance is given by:
\begin{align}
    \frac{\rho_{\chi,0}}{\rho_{\chi,0}^{\mathrm{obs}}}&=8.95\times10^{-6}\frac{T_e^3m_\chi}{M_P^3}\frac{2\pi^2g_{*,S}(t_0)T_0^3}{45\rho_{\chi,0}^{\mathrm{obs}}}\nonumber\\
    &=5.05\times10^{22}\frac{T_e^3m_\chi}{M_P^4},
\end{align}
where $g_{*,S}(t_0)=3.91$ and $T_0=2.73\mathrm K=9.65\times10^{-32}M_P$ is taken.

\begin{figure}[!t]
    \centering\includegraphics[width=0.49\textwidth]{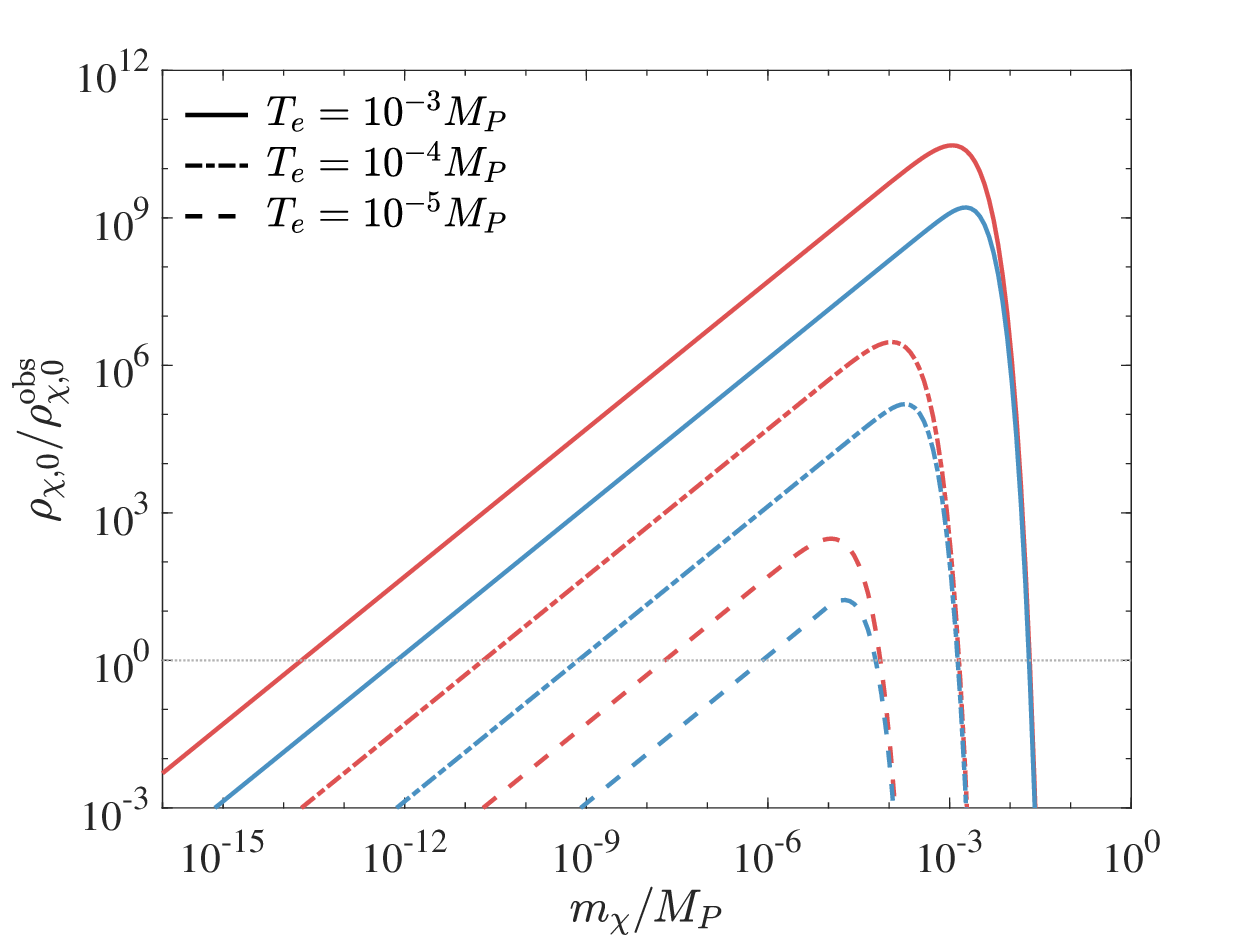}
    \caption{Relic abundance of scalar DM produced by the gravitational annihilation of SM particles (red curves) and inflatons (blue curves). We have assumed $m_\phi\ll T_e$.\label{fig4}}
\end{figure}

The numerical results for several specific WI models with nearly constant temperature before the end of WI are shown as the red curves in Fig.~\ref{fig4}, which roughly conforms to:
\begin{align}\label{absm}
    \frac{\rho_{\chi,0}}{\rho_{\chi,0}^{\mathrm{obs}}}\simeq\left(4.7\pm0.3\right)\times10^{22}\frac{T_e^3m_\chi}{M_P^4}\exp\left[-c_1\left(\frac{m_\chi}{T_e}\right)^{c_2}\right],
\end{align}
where $c_1=0.90$, $c_2=1.17$ for the best fitting. Manifestly, the coefficient is almost consistent with our above analytical derivation, indicating the effectiveness of our simple approximations. Furthermore, this result is similar to that of the SI scenario with $T_e$ replaced by $T_\mathrm{reh}$ \cite{Tang:2017hvq}, except for a slight difference in the coefficient.

We can constrain the mass of DM according to the requirement of $\rho_{\chi,0}\leq\rho_{\chi,0}^{\mathrm{obs}}$. The feasible mass is:
\begin{align}
    &m_\chi\leq2.1\times10^{-23}M_P^4/T_e^3,\\
    \text{or }&m_\chi\gtrsim T_e\left[\frac{4}{c_1}\ln\left(\frac{0.76T_e}{10^{-6}M_P}\right)\right]^{\frac{1}{c_2}}.
\end{align}
Consequently, if $T_e>2.4\times10^{-6}M_P$, the DM with proper mass can be sufficiently produced, thereby constitutes the entirety of the DM abundance. And for the maximum possible temperature ($\sim10^{-3}M_P$), the maximum mass of purely gravitational DM candidate can be up to $10^{-2}M_P$. For fermion, vector, and non-minimally coupled scalar DM candidates, the conclusions are analogous, with only a fine-tuning in coefficients. 

\subsection{Annihilation of inflatons}
We now evaluate inflatons as the initial state in the context of annihilation processes. In the SI models, inflatons are absolutely cold during inflation, characterized by the distribution:
\begin{align}
    f=n_\phi(2\pi^2)\delta^3(\vec p).
\end{align}
Then the interaction term in the Boltzmann equation can be derived as \cite{Clery:2021bwz}:
\begin{align}\label{RphiSI}
    \mathcal R_{\phi\rightarrow\chi}^{\mathrm{SI}}=\frac{\rho_\phi^2}{128\pi M_P^4}\left(1+\frac{m_\chi^2}{2m_\phi^2}\right)^2\sqrt{1-\frac{m_\chi^2}{m_\phi^2}},
\end{align}
for $\phi^2$ potential and scalar DM with $m_\chi<m_\phi$, which is independent of temperature. The relic abundance of DM is estimated as:
\begin{align}
    \frac{\rho_{\chi,0}}{\rho_{\chi,0}^{\mathrm{obs}}}\simeq3\times10^{25}\frac{T_\mathrm{reh}m_\phi m_\chi}{M_P^3},
\end{align}
for $m_\chi\ll m_\phi$. Normally, $m_\phi\gg T_\mathrm{reh}^2/M_P$ is satisfied, hence the DM production from the annihilation of inflatons is more efficient than that of the SM particles in the SI scenario.

The situation in the WI models, however, is quite different. Due to the strong dissipation effect with radiation, the inflatons are generally in a quasi-thermal state \cite{Bartrum:2013fia}. This means the distribution of inflatons should approach the Bose–Einstein distribution. Given that $m_\phi\ll T_e$ is usually satisfied at the end of WI ($T_e\sim\sqrt{0.1m_\phi\phi_e}$), it can also be approximated as the Maxwell–Boltzmann distribution, which allows for the calculation of DM yield utilizing Eq.~(\ref{Y}). In this situation, the interaction term is derived as:
\begin{align}\label{RphiWI}
    \mathcal R_{\phi\rightarrow\chi}^{\mathrm{WI}}=\frac{1}{40\pi^5}\frac{T^8}{M_P^4},
\end{align}
for $m_\phi^2/s\rightarrow0$ and $m_\chi^2/s\rightarrow0$. Then the DM relic abundance is depressed by a factor of $4/149$ compared with Eq.~(\ref{absm}), shown as the blue curves in Fig.~\ref{fig4}.

In contrast to the annihilation of inflatons in the SI case with the same $m_\phi$, the WI case exhibit two significant differences. Firstly, the DM with $m_\chi>m_\phi$ can be produced in the WI case due to the quasi-thermal state of inflatons, while it is prohibited in the SI case. Secondly, if $\phi_e\sim 0.1-1M_P$ is estimated at the end of WI, we will find that the yield of DM with $m_\chi\ll m_\phi$ in the WI case is roughly $10^{-6}$ lower than that in the SI case with $T_\mathrm{reh}=T_e$. It reveals the enormously higher annihilation efficiency of static inflatons than moving inflatons. 

To understand this, we take $\rho_r\simeq\rho_\phi$ at the end of WI and rewrite Eq.~(\ref{RphiWI}) as:
\begin{align}
    \mathcal R_{\phi\rightarrow\chi}^{\mathrm{WI}}\simeq\frac{30^2\rho_\phi^2}{40\pi^7g_*^2M_P^4}\simeq \frac{\rho_\phi^2}{40\pi g_*^2M_P^4}.
\end{align}
Then comparing it with Eq.~(\ref{RphiSI}), we observe that the interaction term for WI is suppressed by a factor of $1/g_*^2\sim 10^{-4}$. The suppression may be attributed to the reduced number density of inflatons in WI. Due to the quasi-thermal state of inflatons in WI, the number density can be estimated as $n_\phi\propto T^3\sim\rho_\phi/T$ with $T\gg m_\phi$. Conversely, in the SI scenario, it is $n_\phi=\rho_\phi/m_\phi$. Therefore, for the same energy density $\rho_\phi$, WI results in a significantly reduced inflaton number density compared to SI, thereby causing a diminished annihilation rate. Actually, $\rho_\phi$ at the end of WI typically takes on a lower value than that in SI, given that it usually holds that $\phi_e^{\mathrm{WI}}<\phi_e^{\mathrm{SI}}$ for the same inflationary potential. As a result, the number density $n_\phi$ in WI and the interaction term $R_{\phi\rightarrow\chi}^{\mathrm{WI}}$ are further suppressed by several orders of magnitude, so does the DM yield.

\subsection{Contribution of CGPP mechanism}
We now consider the contribution of CGPP mechanism. It refers to the particle production from the vacuum of an
expanding universe as we mentioned in the first section. The calculation of CGPP mechanism for a purely gravitational scalar DM candidate $\chi$ should be started with the following action:
\begin{align}\label{SDM}
    S=\int\mathrm d^4x&\sqrt{-g}\left[\frac{M_P^2}{2}R-\frac{\xi}{2}\chi^2R\right.\nonumber\\
    &\left.-\frac{1}{2}\partial_\mu\chi \partial^\mu\chi-\frac{1}{2}m_\chi^2\chi^2+\mathcal L_{\phi,\mathrm{SM}}\right],
\end{align}
where $\mathcal L_{\phi,\mathrm{SM}}$ denotes the part of inflaton and SM particles, the parameter $\xi$ measures the coupling between $\chi$ and gravity. With the conformal time $\mathrm d\tau\equiv \mathrm dt/a$ and the following Fourier mode:
\begin{align}\label{chiexpand}
    a\chi(\tau,\vec x)=\int\frac{\mathrm d^3k}{(2\pi)^3}\left[\hat b_{\vec k}\chi_k(\tau)+\hat b_{-\vec k}^\dag \chi_k^*(\tau)\right]e^{i\vec k\cdot\vec x},
\end{align}
the equation of motion for the mode function $\chi_k(\tau)$ can be written as:
\begin{align}\label{EOM}
    \chi_k''+\omega_k^2\chi_k=0,
\end{align}
where the frequency term
\begin{align}\label{w2}
    \omega_k^2=k^2+a^2m_\chi^2-\left(1-6\xi\right)\frac{a''}{a}.
\end{align}
According to the Friedmann equation of the second kind, we have:
\begin{align}
    \frac{a''}{a}=\frac{a^2}{6M_P^2}\left(\rho-3p\right)=\frac{4a^2V-\phi'^2}{6M_P^2}.
\end{align}
It implies that the contribution of radiation to the frequency term is offset, resulting in no difference in the equation of motion between WI and SI scenarios. Nonetheless, due to the strong dissipation effect, the kinetic term of inflaton field is far smaller than the potential even after the end of WI. Therefore, compared to the SI scenario, the CGPP of DM in WI is free of the influence of inflaton oscillations.

With a specific WI background, the mode functions of DM field can be numerically solved from Eq.~(\ref{EOM}). Then the renormalized energy density at $\tau\rightarrow+\infty$ limit can be calculated by:
\begin{align}\label{rho}
    \rho_{\chi}=\int \frac{k^2 \mathrm dk}{4\pi^2a^4}\left[\left|\chi_{k}'\right|^2+\left(k^2+a^2 m_{\chi}^2\right)\left|\chi_{k}\right|^2-\omega_k\right].
\end{align}
For the case of $\xi\leq1/6$ and $m_\chi\gtrsim\sqrt{2(1-6\xi)H_\mathrm{inf}}$, where $H_\mathrm{inf}$ denotes the typical value of Hubble parameter long before the end of inflation, the frequency term is always positive, resulting in a finite integral in Eq.~(\ref{rho}) with main contribution coming from the interval around $k/a_e\sim\sqrt{H_em_\chi}$. To reiterate, the subscript $e$ denotes the end of WI. However, when $m_\chi<\sqrt{2(1-6\xi)H_\mathrm{inf}}$ and $\xi<1/6$, tachyonic modes with negative frequency term emerge for small $k$. They lead to the infrared divergence of the integral, and thus necessitate a cut-off as a lower bound for the calculation of integral. A suitably motivated cut-off is $k>a_0H_0$, as modes with $k$ below this threshold are outside the horizon today and, in general, do not contribute to the DM energy density.

\begin{figure}[!t]
    \centering\includegraphics[width=0.49\textwidth]{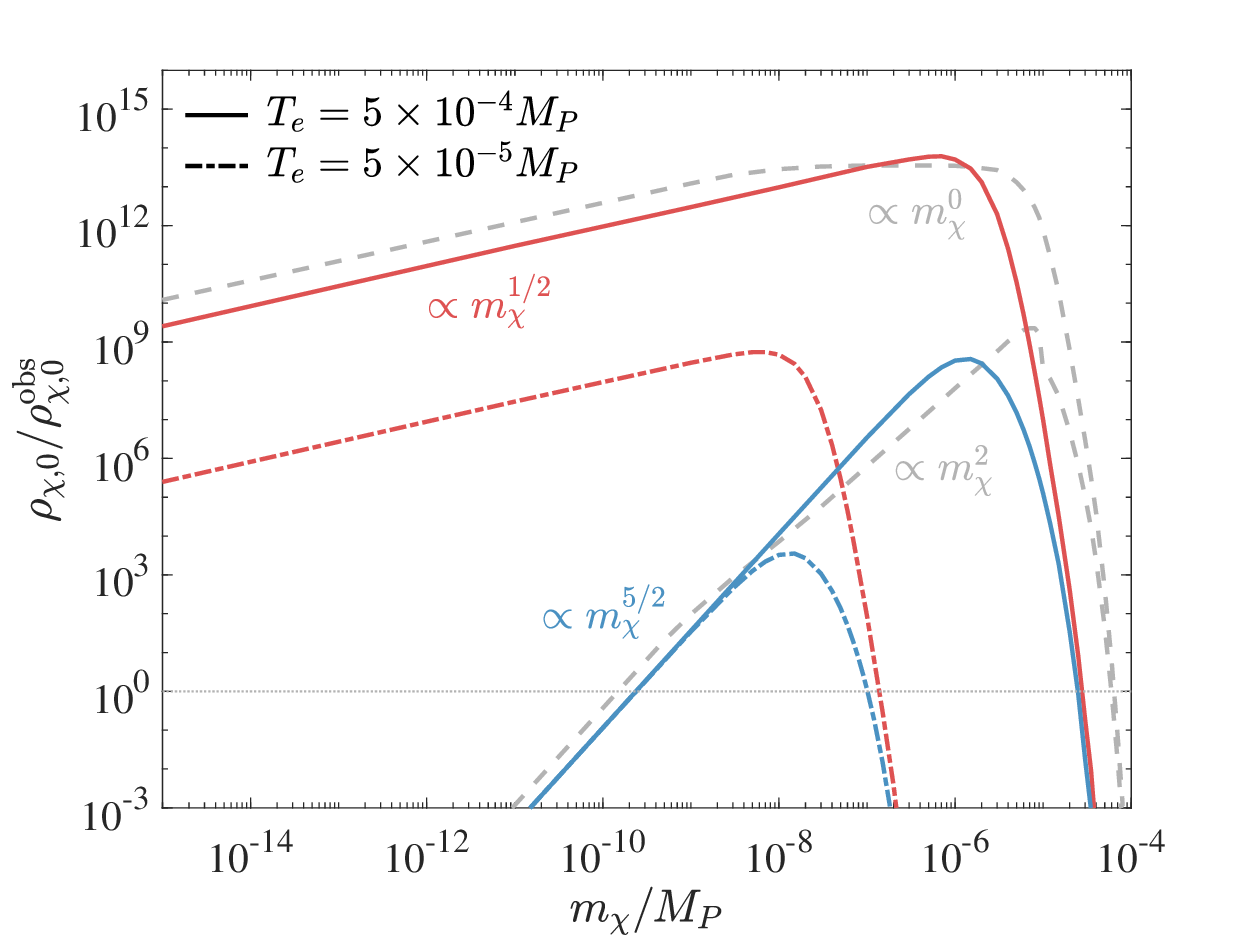}
    \caption{Relic abundance of scalar DM with minimal (red curves) and conformal (blue curves) coupling via CGPP mechanism in WI. The inflationary background is served by a $\phi^2$ WI model. We set $m_\phi=10^{-5}(10^{-7})M_P$, $T_e=5\times10^{-4}(5\times10^{-5})M_P$, and $H_e=10^{-6}(10^{-8})M_P$ for the solid(dashed) curves. For comparison, the gray dashed curves depict DM relic abundance in the $\phi^2$ SI scenario with $m_\phi=10^{-5}M_P$ and $T_\mathrm{reh}=3\times10^{-5}M_P$.\label{fig5}}
\end{figure}

We adopt a $\phi^2$ WI model with mass parameters $m_\phi=10^{-5}M_P$ and $10^{-7}M_P$. By adjusting the parameters in Eq.~(\ref{Upsl}), we obtain WI backgrounds with $T_e=5\times10^{-4}M_P$, $H_e=10^{-6}M_P$ and $T_e=5\times10^{-5}M_P$, $H_e=10^{-8}M_P$, respectively. The CGPP of DM with minimal coupling ($\xi=0$) and conformal coupling ($\xi=1/6$) is numerically investigated. The abundance-mass correlations are shown in Fig.~\ref{fig5}. These results are actually indicative of the general characteristics of a wide range of WI models, although they are based on a specific model.

We first focus on the $m_\chi<H_e$ case, the relic abundance roughly confirms to:
\begin{align}\label{abmini}
    &\frac{\rho_{\chi,0}}{\rho_{\chi,0}^{\mathrm{obs}}}\simeq1.5\times10^{30}\frac{T_e^4m_\chi^{1/2}}{M_P^{9/2}},~~~\xi=0,\\
    &\frac{\rho_{\chi,0}}{\rho_{\chi,0}^{\mathrm{obs}}}\simeq1.1\times10^{24}\frac{m_\chi^{5/2}}{M_P^{5/2}},~~~\xi=1/6.\label{abconf}
\end{align}
It manifests a feasible DM mass range of:
\begin{align}
    &m_\chi\leq4.5\times10^{-61}\frac{M_P^9}{T_e^8},~~~\xi=0,~T_e>10^{-6}M_P,\\
    &m_\chi\leq2.4\times10^{-10}M_P,~~~\xi=1/6,~T_e>10^{-5}M_P.
\end{align}
Here we present a threshold for $T_e$. If WI takes place with $T_e$ below this threshold, there will not be an overproduction of DM for any mass. By comparison, we calculate the DM relic abundance in the $\phi^2$ SI model with $m_\phi=10^{-5}M_P$ and $T_\mathrm{reh}=3\times10^{-5}M_P$, shown as the gray dashed curves in Fig.~\ref{fig5}. It has a correlation of
\begin{align}
    &\rho_{\chi,0}\propto m_\chi^{0},~~~H_\mathrm{reh}\lesssim m_\chi\lesssim H_e,\\
    &\rho_{\chi,0}\propto m_\chi^{1/2},~~~~~m_\chi\ll H_\mathrm{reh},
\end{align}
for minimal coupling, and
\begin{align}
    &\rho_{\chi,0}\propto m_\chi^2,~~~H_\mathrm{reh}\lesssim m_\chi\lesssim H_e,\\
    &\rho_{\chi,0}\propto m_\chi^{5/2},~~~~~m_\chi\ll H_\mathrm{reh},
\end{align}
for conformal coupling, where $H_\mathrm{reh}$ is the Hubble parameter when reheating ends. To a certain extent, WI is similar to the SI with instantaneous reheating, corresponding to $H_\mathrm{reh}=H_e$.

For the $m_\chi>H_e$ case, apparently, a little increase in mass leads to a rapid decrease in the relic abundance. The super-Hubble-mass DM with $m_\chi\sim10-30H_e$ can be sufficiently produced and constitutes the entirety of observed DM abundance. This is similar to the SI case. Nevertheless, for the same inflationary potential, WI generally has lower $\phi_e$ or $H_e$ than SI. It actually narrows the DM mass range of overproduction compared with SI, thereby expanding the feasible DM mass range. This is clearly demonstrated in Fig.~\ref{fig5}.

\section{Observational constraints}
\begin{figure}[!t]
    \centering\includegraphics[width=0.49\textwidth]{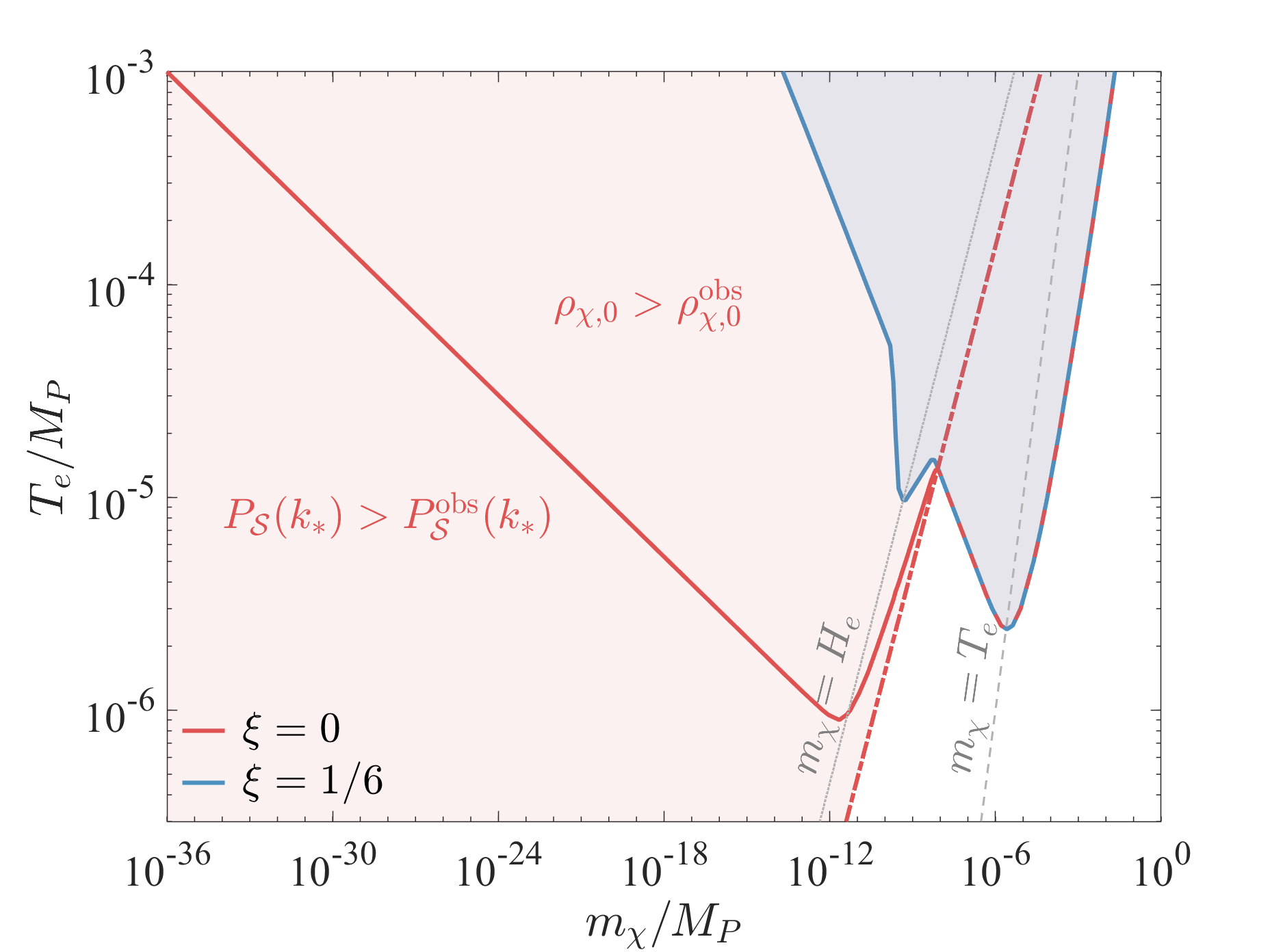}
    \caption{Observational constraints on the temperature $T_e$ and DM mass $m_\chi$. The red and blue areas are the excluded parameter spaces respectively for the minimal coupling ($\xi=0$) and the conformal coupling ($\xi=1/6$). The red and blue curves refer to the correlation between temperature $T_e$ and DM mass $m_\chi$ corresponding to $\rho_{\chi,0}=\rho_{\chi,0}^{\mathrm{obs}}$. The impact of coupling $\xi$ on the gravitational annihilation channel is ignored. The red dashed line refers to $P_\mathcal{S}(k_*)=P_\mathcal{S}^{\mathrm{obs}}(k_*)$. The gray dotted(dashed) line marks $m_\chi=H_e(T_e)$. \label{fig6}}
\end{figure}

In this section, we give the observational constraints for the parameter space. First, we depict the abundance constraints with all possible DM production channels taking into account, including the CGPP mechanism alongside the gravitational annihilation of SM particles and inflatons, as shown in Fig.~\ref{fig6}. The red and blue curves depict the cases where $\rho_{\chi,0}=\rho_{\chi,0}^{\mathrm{obs}}$ respectively for the minimal and conformal coupling, and the areas enclosed by them represent the excluded overproduction situations. Here we ignore the impact of coupling $\xi$ on the gravitational annihilation channel. This approximation is advisable for $|\xi|\ll1$.

Another constraint comes from the isocurvature perturbations \cite{Chung:2004nh}. Observations by the Planck Collaboration restrict the primordial isocurvature fraction, $\beta_\mathrm{iso}\equiv P_{\mathcal S}(k_*)/\left(P_{\mathcal S}(k_*)+P_{\mathcal R}(k_*)\right)<0.038$, at the CMB pivot scale $k_*$, leading to an upper limit of the isocurvature power spectrum, $P_{\mathcal S}^{\mathrm{obs}}(k_*)<8.3\times10^{-11}$. Isocurvature perturbations can be induced by the CGPP of DM. For purely gravitational scalar DM, we define a parameter $\alpha\equiv4\xi+m_\chi^2/\left(3H_*^2\right)$, the induced isocurvature power spectrum at $k_*$ then can be estimated as follows \cite{Garcia:2023awt,Garcia:2023qab}:
\begin{gather}
    P_{\mathcal S}(k_*)\simeq\frac{4\left(N_{\mathrm{tot}}-N_*\right)\alpha^2e^{-4N_*\alpha}}{\left(1-e^{-2N_\mathrm{tot}\alpha}\right)^2},~~~\alpha\lesssim3/4,\\
    P_{\mathcal S}(k_*)\simeq\frac{3}{2}e^{-3N_*},~~~\alpha\gtrsim3/4,
\end{gather}
where $N_*\sim60$ denotes the $e$-folding number between the $k_*$ crossing and the end of inflation. $N_\mathrm{tot}$ denotes the total $e$-folding number during the inflationary phase, and its value is dependent on the choice of the infrared cut-off of $k$. Consistent with previous discussions, we set the cut-off as $k>a_0H_0$, which leads to $N_\mathrm{tot}\sim65$. Combining the observational constraint and the above equations, we restrict that $\alpha>0.089$. For scalar DM with conformal coupling, this condition is inherently fulfilled, thereby eliminating any constraints on its mass. However, for the minimally coupled scalar DM, this condition implies that $m_\chi>0.52H_*\sim m_\phi$, which means that only DM with a super-Hubble scale mass is allowed in this case. This constraint is depicted by the red dashed line in Fig.~\ref{fig6}. The mass ranges of DM with sufficient density are limited to $10^{-8}-10^{-2}M_P$ for minimal coupling and $10^{-14}-10^{-2}M_P$ for conformal coupling.

Measurements of Lyman-$\alpha$ forest also provide constraints on the mass of purely gravitational DM by imposing limits on the equation of state of DM \cite{Ballesteros:2020adh,Garcia:2022vwm}. According to the result in \cite{Garcia:2022vwm}, a mass limit of $m_\chi>\mathcal O(\mathrm{eV})$ is established for a large enough reheating temperature in the SI scenario. This result is also applicable for the WI scenario. However, for the minimal coupling case, this constraint is considerably less stringent compared to that associated with isocurvature perturbations. And for conformal coupling, DM with a mass less than $\mathcal O(\mathrm{eV})$ has a negligible yield, rendering them incapable of being the principal component of DM that influences Lyman-$\alpha$ forest observations. Therefore, in these two cases, Lyman-$\alpha$ forest measurements actually do not provide stricter constraints than our previous discussions.

\section{Conclusion}
We have investigated the purely gravitational DM production within the WI background. Three possible gravitational channels for DM production are explored, and exhibit distinctive features compared to the SI scenario. For the freeze-in channel via the SM particles annihilation, we derive the interaction term in the Boltzmann equation and calculate the DM yield. The analytical approach is utilized for the light DM case, which is consistent with the numerical results. For the inflaton annihilation channel, we obtain markedly different results from the SI scenario. Due to the quasi-thermal state of inflaton, the channel of producing super-inflaton-mass DM is open in WI, while the sub-inflaton-mass case is suppressed compared to the SI scenario. In the context of the CGPP channel, we identify the temperature threshold for sufficient DM production. We also show a correlation of $\rho_{\chi,0}\propto T_e^4m_\chi^{1/2}$ for minimally coupled DM and $\rho_{\chi,0}\propto m_\chi^{5/2}$ for conformally coupled DM, indicating differences from the SI scenario. Furthermore, the constraints on the DM mass $m_\chi$ and the temperature $T_e$ at the end of WI are explored. We demonstrate the feasible parameter space limited by the observations of DM abundance and isocurvature perturbations. The results of this study can be extended to purely gravitational fermion and vector DM, as they are similar to the scalar DM with conformal and minimal coupling, respectively.

The results of this paper may help distinguish between WI and SI if combined with the future observations of primordial gravitational waves. The energy scale of inflation at the CMB pivot scale is given by $V_*=\frac{3\pi^2}{2}r\Delta_R$. If the tensor-to-scalar ratio $r$ can be measured, we can confirm $V_*$ and estimate the energy scale when inflation ends, $V_e$. In WI, this means the temperature $T_e$ is confirmed. And the possible mass of purely gravitational DM with sufficient abundance can be read from the mass-temperature relationship we derived (shown as Fig.~\ref{fig6}). In SI, however, the production of DM depends not only on $V_e$ but also on $T_\mathrm{reh}$, so confirming $V_e$ cannot restrict the DM mass with sufficient abundance. In this way, if purely gravitational DM constitutes all of the DM, one can distinguish the WI and SI by the measurement of DM mass (assuming there is a way to measure it in the future). If the measured mass align with the mass-temperature relationship of WI, it would suggest that the inflation scenario might be WI. Otherwise, if the observed mass does not align, it is more likely to be SI. The above method is very idealized, but feasible in principle.

To summarize, this work establishes WI as a possible scenario for purely gravitational DM production, offering novel solutions to the DM origins. By bridging thermal history with gravitational interactions, it opens new avenues for exploring DM and inflationary dynamics.

\appendix*
\section{Derivation of the Boltzmann equation}
Now we give a detailed derivation for Eq.~(\ref{RT}). We first introduce the Boltzmann equation for $1+2\leftrightarrow3+4$ scattering process in an expanding universe:
\begin{align}\label{blzm1}
    &\frac{1}{a^3}\frac{\mathrm d\left(n_3a^3\right)}{\mathrm dt}=\int\prod^4_{j=1}\frac{\mathrm d^3\vec p_j}{\left(2\pi\right)^3 2E_j}\nonumber\\
    &\times\left(2\pi\right)^4\delta^4\left(p_1+p_2-p_3-p_4\right)\sum_{\mathrm{pol}}\left|\mathcal M\right|^2\nonumber\\
    &\times\left[f_1f_2\left(1\pm f_3\right)\left(1\pm f_4\right)-f_3f_4\left(1\pm f_1\right)\left(1\pm f_2\right)\right],
\end{align}
here $n_3$ is the number density of final particles, namely DM. And
\begin{align}
    f=\frac{1}{\exp\left[\left(E-\mu\right)/T\right]\pm1},
\end{align}
is the distribution function, where $+1$ corresponds to fermion and $-1$ corresponds to boson. For the process we consider, the chemical potential $\mu_{1,2}\gg\mu_{3,4}$, and the main contribution comes from particles with $E-\mu\gg T$, therefore $f\ll1$ and $f_{1,2}\gg f_{3,4}$. Then Eq.~(\ref{blzm1}) can be simplified as:
\begin{align}
    \frac{1}{a^3}&\frac{\mathrm d\left(n_3a^3\right)}{\mathrm dt}=\int\prod^4_{j=1}\frac{\mathrm d^3\vec p_j}{\left(2\pi\right)^3 2E_j}\nonumber\\
    &\times\left(2\pi\right)^4\delta^4\left(p_1+p_2-p_3-p_4\right)\sum_{\mathrm{pol}}\left|\mathcal M\right|^2f_1f_2.
\end{align}
Actually, $E-\mu\gg T$ implies that the distribution can be approximated as the Maxwell–Boltzmann distribution, $f=e^{-E/T}$. This will facilitate an analytical expression for the final result.

Integrating over the momentum of the final particles in the center-of-mass frame, we derive:
\begin{align}
    &\int\frac{\mathrm d^3\vec p_3}{2E_3}\frac{\mathrm d^3\vec p_4}{2E_4}\frac{1}{\left(2\pi\right)^2}\delta^4\left(p_1+p_2-p_3-p_4\right)\sum_{\mathrm{pol}}\left|\mathcal M\right|^2\nonumber\\
    =&\int\frac{\mathrm d^3\vec p_3\mathrm d^3\vec p_4}{\left(2\pi\right)^2s}\delta\left(\sqrt{s}-E_3-E_4\right)\delta^3\left(-\vec p_3-\vec p_4\right)\sum_{\mathrm{pol}}\left|\mathcal M\right|^2\nonumber\\
    =&\int\frac{1}{\left(2\pi\right)^2s}\frac{\mathrm d^3\vec p_3}{\mathrm d\sqrt{s}}\sum_{\mathrm{pol}}\left|\mathcal M\right|^2=\int\frac{\left|\vec p_3\right|^2}{2\pi s}\frac{\mathrm d\left|\vec p_3\right|}{\mathrm d\sqrt{s}}\mathcal A\nonumber\\
    =&~4g_1g_2\sigma\sqrt{\left(p_1\cdot p_2\right)^2-m_1^2m_2^2},
\end{align}
Then the Boltzmann equation reduce to:
\begin{align}\label{blzm2}
    \frac{1}{a^3}\frac{\mathrm d}{\mathrm dt}\left(n_3a^3\right)=&\int\frac{f_1g_1\mathrm d^3\vec p_1}{\left(2\pi\right)^3 E_1}\frac{f_2g_2\mathrm d^3\vec p_2}{\left(2\pi\right)^3 E_2}\nonumber\\
    &\times\sigma\sqrt{\left(p_1\cdot p_2\right)^2-m_1^2m_2^2}.
\end{align}
According to the procedure in \cite{Gondolo:1990dk,Edsjo:1997bg}, this equation can be further reduced with:
\begin{align}\label{volume}
    \mathrm d^3\vec p_1\mathrm d^3\vec p_2&=8\pi^2|\vec p_1|E_1\mathrm dE_1|\vec p_2|E_2\mathrm dE_2\mathrm d\cos\theta\nonumber\\
    &=2\pi^2E_1E_2\mathrm dE_+\mathrm dE_{-}\mathrm ds,
\end{align}
where $\theta$ is the angle between $\vec p_1$ and $\vec p_2$, and
\begin{gather}
    E_+\equiv E_1+E_2,~~~E_{-}\equiv E_1-E_2,\\
    s=m_1^2+m_2^2+2E_1E_2-2|\vec p_1||\vec p_2|\cos\theta.
\end{gather}
We assume that $m_1=m_2=m_i$, $m_3=m_4=m_f$, then the integration range is given by:
\begin{gather}
    s\geq\mathrm{max}\left(4m_{i}^2,4m_f^2\right),\\
    E_+\geq\sqrt s,\\
    |E_{-}|\leq\sqrt{1-\frac{4m_i^2}{s}}\sqrt{E_+^2-s}.
\end{gather}
Substituting Eq.~(\ref{volume}) into Eq.~(\ref{blzm2}), and integrating over $E_-$ and $E_+$ in sequence, we finally derive:
\begin{align}
    \frac{1}{a^3}\frac{\mathrm d}{\mathrm dt}\left(n_3a^3\right)=~&\frac{g_i^2}{32\pi^4}\int\mathrm ds\int\mathrm dE_+e^{-E_+/T}\int\mathrm dE_{-}\nonumber\\
    &\times\sigma\sqrt{\left(p_1\cdot p_2\right)^2-m_i^4}.\nonumber\\
    =~&\frac{g_i^2}{32\pi^4}\int\mathrm ds\int\mathrm dE_+e^{-E_+/T}\nonumber\\
    &\times\sigma\left(s-4m_i^2\right)\sqrt{E_+^2-s}\nonumber\\
    =~&\frac{Tg_i^2}{32\pi^4}\int\mathrm ds\sigma\sqrt{s}\left(s-4m_i^2\right)K_1\left(\frac{\sqrt{s}}{T}\right)\nonumber\\
    \equiv~&\mathcal R(T).
\end{align}
The Maxwell–Boltzmann distribution is adopted in the above derivation.

\bibliography{ref.bib}

\end{document}